# Quantum percolation phase transition and magneto-electric dipole glass in hexagonal ferrites

S. E. Rowley[1,2,3], T. Vojta[4], A. T. Jones[1,5], W. Guo[1], J. Oliveira[2], F. D. Morrison[6], N. Lindfield[6], E. Baggio Saitovitch[2], B. E. Watts[7] and J. F. Scott[8]

[1]Cavendish Laboratory, Cambridge University, J. J. Thomson Avenue, Cambridge, CB3 0HE, United Kingdom

[2]Centro Brasileiro de Pesquisas Físicas, Rua Dr Xavier Sigaud 150, Rio de Janeiro, 22290-180, Brazil

[3]Quantum Materials Laboratory, Cambridge, CB3 9NF, United Kingdom

[4]Physics Department, Missouri University of Science and Technology, Missouri, 65409, United States of America

[5]Physics Department, Lancaster University, Lancaster, LA1 4YB , United Kingdom

[6]School of Chemistry, St. Andrews University, St. Andrews, Fife, KY16 9ST, United Kingdom

[7]IMEM-CNR, Parco Area delle Scienze 37/A, 43124 Parma, Italy

[8]Schools of Chemistry and of Physics and Astronomy, St. Andrews University, St. Andrews, Fife, KY16 9ST, United Kingdom

**Hexagonal ferrites do not only have enormous commercial impact (£2 billion/year in sales) due to applications that include ultra-high density memories, credit card stripes, magnetic bar codes, small motors and low-loss microwave devices, they also have fascinating magnetic and ferroelectric quantum properties at low temperatures. Here we report the results of tuning the magnetic ordering temperature in $PbFe_{12-x}Ga_xO_{19}$ to zero by chemical substitution $x$. The phase transition boundary is found to vary as $T_N \sim (1 - x/x_c)^{2/3}$ with $x_c$ very close to the calculated spin percolation threshold which we determine by Monte Carlo simulations, indicating that the zero-temperature phase transition is geometrically driven. We find that this produces a form of compositionally-tuned, insulating, ferrimagnetic quantum criticality. Close to the zero temperature phase transition we observe the emergence of an electric-dipole glass induced by magneto-electric coupling. The strong frequency behaviour of the glass freezing temperature $T_m$ has a Vogel-Fulcher dependence with $T_m$ finite, or suppressed below zero in the zero frequency limit, depending on composition $x$. These quantum-mechanical**

**properties, along with the multiplicity of low-lying modes near to the zero-temperature phase transition, are likely to greatly extend applications of hexaferrites into the realm of quantum and cryogenic technologies.**

M-type hexagonal ferrites (hexaferrites) including $BaFe_{12}O_{19}$, $SrFe_{12}O_{19}$ and $PbFe_{12}O_{19}$ are popular magnetic materials for their use in a wide range of applications [1, 2]. Moreover, they also have interesting magnetic and ferroelectric properties at low temperatures [3, 4]. Here we study the effects of tuning the magnetic ordering (Néel) temperature all the way to zero resulting in a geometrically driven zero-temperature phase transition of the underlying spin system and the emergence of an electric-dipole glass. These properties are expected to be important for a wide range of advanced quantum and cryogenic applications including, for example, electro-caloric and magneto-caloric refrigeration, and quantum memory devices, as the materials can be readily controlled by magnetic fields and voltage gates.

$BaFe_{12}O_{19}$, $SrFe_{12}O_{19}$ and $PbFe_{12}O_{19}$ crystalise in the magnetoplumbite structure and are Lieb-Mattis [5] ferrimagnets with Néel temperatures of approximately $T_N \approx 720K$ and saturated magnetisations in the low temperature limit of $20\mu_B$ per double formula unit [6]. The crystal structure can be seen in the right inset to Fig. 1a which shows a double unit-cell. The underlying spin structure comprises collinear anti-ferromagnetic order below $T_N$ with a total of 16 spins pointing up and 8 spins pointing down located on $Fe^{3+}$ sites per double unit-cell resulting in ferrimagnetism. The $Fe^{3+}$ ions, each in the high $S = 5/2$ spin state, are located on five sub-lattices as follows: six spin-up on octahedral sub-lattice (k), one spin-up on octahedral sub-lattice (2a), one spin-up on pseudo-hexahedral sub-lattice (2b), two spin-down on tetrahedral sub-lattice ($4f_{IV}$) and two spin-down on octahedral sub-lattice ($4f_{VI}$) [6]. The M-type hexaferrites are n-type semiconductors [7] with bandgaps of $E_g \approx 0.63$ eV and rather heavy electrons and holes: $m$(light e) = 5.4 $m_e$; $m$(heavy e) = 15.9 $m_e$; $m$(light h) = 10.2 $m_e$; $m$(heavy h) = 36.2 $m_e$ and highly anisotropic conductivity. For electric fields applied normal to the c-axis, the electrical conductivity is circa fifty times greater than along *c*. An example of the hexahedral (bi-pyramid) sites is shown in the right inset to Fig. 1a where its faces have been shaded in grey. All of the bi-pyramids comprise a single $Fe^{3+}$ ion at the centre surrounded by five $O^{2-}$ ions at the corners. The vibration of these positive ions within their negatively charged oxygen enclosures is along the *c* direction and generates polar transverse-optic $A_{2u}$-symmetry phonon modes. The lowest of these is observed to drop in frequency at long wavelengths ($q = 0$) to 42 cm$^{-1}$ (5.2meV) as *T* approaches zero resulting in an incipient ferroelectric state [8, 9]. This state has been investigated in recent work with evidence of uniaxial ferroelectric quantum critical behaviour along the *c* direction [3] and anti-ferroelectric frustration on the triangular lattice of dipoles in the *a-b* plane [4].

In this paper we study a different part of the phase diagram close to the insulating magnetic zero temperature critical point, achieved by suppressing $T_N$ to zero by randomly substituting Ga ions for the Fe ions in $PbFe_{12-x}Ga_xO_{19}$. As *x* increases, the lattice of spins is diluted as the non-magnetic gallium ions act as quenched spinless impurities. This results in a drop in $T_N$ as determined by Mössbauer and magnetic measurements [6]. The Mössbauer data also indicate that the Ga ions distribute themselves with nearly equal probability in all the available sublattices, at least for not too large *x*. As shown in Fig. 1, by extrapolating the trend to $T = 0$K we find that the critical value of *x* for which $T_N$ goes to zero is $x = x_c \approx 8.6$.

The zero temperature transition between the ferrimagnetic and nonmagnetic ground states as a function of iron concentration can either be geometrically driven or driven by quantum fluctuations. In the first scenario, the transition is a percolation transition. It occurs when the iron concentration falls below the percolation threshold $p_c$ of the lattice of iron sites where *p* is the probability of a site containing an iron atom [related to *x* by $x = 12(1-p)$ ]. Long-range magnetic order is then impossible because the iron atoms form disconnected finite-size clusters. In the second scenario, the zero-temperature phase transition occurs before the iron concentration falls below $p_c$ because the magnetic order is destroyed by quantum fluctuations of the iron spins.

To help distinguishing the two scenarios, we have determined the percolation threshold of the lattice of iron atoms in the M-type hexagonal ferrites by means of computer simulations. This requires knowledge about the connectivity of the iron atoms, i.e., about the exchange interactions between the iron atoms on the five different sub-lattices. The exchange interactions in $BaFe_{12}O_{19}$ were determined from phenomenological fits of experimental sub-lattice magnetization data in Refs. [10] and [11]. More recently, these interactions were also calculated from first principles [12]. Even though the exact values of the interactions differ between these papers, they agree on the basic structure: The four dominating interactions are between the following sub-lattices: 2a-$4f_{IV}$, 2b-$4f_{VI}$, 12k-$4f_{IV}$, and 12k-$4f_{VI}$ (see Fig 1a). These interactions are antiferromagnetic, and they are not frustrated because each couples a spin-up and a spin-down sub-lattice. All other interactions are significantly weaker, and they are frustrated because they couple spins in the same sub-lattice or in different sub-lattices with the same spin direction. The exchange interactions in all the M-type hexagonal ferrites, Pb, Sr and Ba are expected to be very similar. In our percolation simulations we have therefore only included the bonds corresponding to the four dominating (unfrustrated) interactions. The weaker frustrated interactions may become important at dilutions close to the percolation threshold and at low temperatures. Because they are frustrated, they are expected to suppress the ferrimagnetic order compared to a scenario that includes only the leading unfrustrated ones. We have further assumed that all iron sites have the same occupation probability (i.e., the gallium doping is completely random).

To find the percolation threshold for the thus defined lattice of iron atoms, we have implemented a version of the fast Monte Carlo algorithm due to Newman and Ziff [13]. We have studied systems with sizes of up to 200x200x200 double unit cells (192 million Fe sites), averaging over several thousand disorder realizations for each size. The percolation threshold is determined from the onset of a spanning cluster. Extrapolating the results to infinite system size, yields $p_c = 0.2628(5)$ where the number in brackets is the error of the last digit (estimated from the very small statistical error of the data and the robustness of the extrapolation). In the material $PbFe_{12-x}Ga_xO_{19}$, this corresponds to a gallium concentration $x = 8.846(6)$. We note that the percolation threshold for our realistic model of the magnetic interactions in the hexaferrites is also very close to the threshold for a simple three-dimensional hexagonal stacked structure [14, 15]. Fig. 1b shows a projection of the relevant Fe ions into the a-b plane for the parent compound (left) and a percolating magnetic cluster (right) for $x$ close to $x_c$.

The experimentally observed value of $x_c = 8.6$ is very close to that determined above from our model calculations suggesting that the zero temperature phase transition is predominantly geometrically driven by the percolation of magnetic ions through the crystal. The closeness of the measured and calculated values of $x_c$ is further evidence, along with the results of Mössbauer experiments [6] mentioned above, that Ga ions are substituted randomly onto the Fe sites of the parent compound. For $x > x_c$, magnetic long-range order is impossible. This is confirmed by the measured heat capacity of a sample with $x = 9$, shown in Fig. 2a, which does not feature a phase transition down to the lowest measured temperatures. However, due to statistical fluctuations in the distribution of the Ga ions, we expect disconnected Fe-rich clusters of magnetic order to exist within a background rich in the non-magnetic Ga ions, resulting in a paramagnetic Griffiths phase [16, 17]. The weak hysteresis observed in the magnetisation-field curve of the x=9 sample, shown in Fig. 2b, supports this picture as it can be attributed to the contribution to the uniform magnetisation from the ferrimagnetic clusters. The experimental confirmation of the paramagnetic Griffiths phase will require further study. At the percolation threshold, and for $x < x_c$, there is an "infinite percolation cluster" that spans the entire crystal, resulting in a finite value of the global $T_N$ and long-range magnetic order. The point $T = 0K$ and $x = x_c$ can be understood as a multi-critical point (MCP) because it combines the geometrical criticality of the zero-temperature percolation transition (characterised by the percolation critical exponents [14]) and the thermal criticality of the finite-temperature phase boundary (characterised by the usual thermodynamic critical exponents). Despite extensive theoretical work on classical and quantum magnetic percolation phase transitions, there are relatively few experimental examples. Notable examples include work on the square lattice two-dimensional $La_2Cu_{12-z}(Zn,Mg)_zO_4$ system [18] and on transition metal halides [19]. $PbFe_{12-x}Ga_xO_{19}$ is unique in that it is a three-dimensional hexagonal magneto-electric system that can be successfully tuned up to and beyond the percolation threshold with a novel phase transition boundary $T_N(x)$ as discussed below. The fact that the measured value of $x_c$ is a little less than that determined from calculations could be due to a small degree of Ga clustering or alternatively due to effects of quantum fluctuations arising from the sub-dominant frustrated magnetic interactions referred to above.

As shown in Fig. 1 we find that the shape of the phase transition boundary follows a striking $T_N \sim (1 - x/x_c)^{2/3}$ dependence over the entire concentration range from $x = 0$ to $x = x_c$. Where does this power law come from? As the phase boundary starts at high temperatures ($T_N(0) = 720K$), one might expect $T_N(x)$ to follow the form predicted by classical percolation theory, $\mu(T_N) \sim (x_c - x)^\phi$ where $\mu(T)$ is the appropriate spin Hamiltonian temperature scaling function. $\mu(T) \sim e^{-2J/k_B T}$ for a system with Ising symmetry and $\mu(T) \sim T$ for a system with continuous (e.g. Heisenberg) symmetry and $\phi$ is a crossover exponent usually defined as the ratio of percolation and thermal correlation length critical exponents $\phi = \nu_p/\nu_T$ [14, 19-21]. Over the range of temperatures and chemical compositions tested so far, our measured $T_N(x)$ curve is quite different from the predictions of these classical models in which usually $\phi \geq 1$.

Alternatively, the shape of the phase boundary may be governed the zero-temperature quantum phase transition occurring as the composition $x$ is tuned through $x_c$. Quantum phase transitions are subtly different from the more familiar classical phase transitions occurring as a function of temperature at high temperatures. In the present case, the zero entropy state at $T = 0$ K (or at sufficiently low temperatures for the third law of thermodynamics to apply) of long-range ferrimagnetic order for $x < x_c$, is transformed into a paramagnetic state with no conventional magnetic ordering for $x > x_c$. This zero temperature state still has zero entropy (assuming a non-degenerate ground state). Magnetic quantum phase transitions are a highly active area of research. Depending on specific material details (precise lattice geometry) and dimensionality, they can lead to a rich tapestry of exotic phases such as dimer states, valence bond solids, spin glasses, quantum spin liquids and topological entities, such as spin spiral states and others [22-26]. In the presence of quenched disorder, quantum phase transitions can give rise to smeared phase transitions as well as to the above-mentioned Griffiths phases [16] that are characterized by singular low-temperature thermodynamic functions with gapless excitations over a range of tuning parameter variables [17, 25, 27, 28].

In 'clean' quantum critical systems, where the interactions are tuned, for example, by lattice density or magnetic field, the effects of quantum criticality can often be felt over a wide range of temperatures and tuning parameters below a temperature scale $T^*$ set for example by the spectrum of magnetic excitations ($T^* = \hbar\omega^*/k_B$). In contrast to classical critical points, thermodynamic properties near quantum critical points are affected by fluctuations of the order-parameter field in space and time. This implies that thermodynamic quantities are functions of the dynamical exponent $z$ characterizing the spectrum $\Omega(q) \sim q^z$ of modes close to the critical point where $q$ is the wavevector. For insulators where the modes are typically propagating (heavily under-damped) $\Omega(q)$ is a frequency-wavevector dispersion of the normal modes, whereas for metals where the modes are typically dissipative (heavily over-damped) $\Omega(q)$ is a relaxation rate spectrum. If $d + z \geq 4$ (where $d$ is the dimension of space, thermodynamic quantities in the quantum critical regime may be calculated by the one-loop (Hartree) approximation used in renormalization group models and self-consistent-field models of quantum criticality[23, 29-41]. Close to the quantum critical point, this yields the magnetic susceptibility $\chi \sim 1/T^{\gamma_T}$ where the (thermal) critical exponent $\gamma_T = (d + z - 2)/z$. The critical temperature is found to vary as $T_c \sim (1 - g/g_c)^{1/\gamma_T}$ where $g$ is the (non-disorder inducing) quantum tuning parameter. We note that the value 2/3 of the phase transition boundary exponent found in Fig. 1 would be consistent with a dynamical exponent $z = 2$ and $d = 3$. Such an exponent $z$ in a magnetic insulator may arise for example from the dynamics of spin precession [23, 25, 42].

However, the situation is different in the presence of strong disorder as introduced, for example by the dilution of the magnetic lattice. According to the Harris criterion [43], disorder is typically a relevant perturbation at a quantum phase transition and therefore destabilizes the clean critical behavior. For the specific case of magnetic percolation quantum phase transitions, theories predict that thermodynamic functions either depend on a new dynamical exponent $z'$ defined in terms of the fractal dimension $D_f$ of the percolation transition [14] and the dynamical exponent $z$ of the clean quantum phase transition [27, 28] or they show even more unconventional activated scaling behavior[20]. However, in both cases, the phase boundary is predicted to follow the classical behavior discussed above, in disagreement with our observation.

The origin of the unusual phase transition boundary with a 2/3 power law may be due to the interaction of magnetic and ferroelectric degrees of freedom, or the combined effects of quantum fluctuations (arising from the frustrated magnetic interactions referred to above) and those of the geometrically-driven percolation transition. In a future study the relative importance of these effects may be separated by employing further tuning parameter such as pressure or field in addition to chemical composition.

We now turn to the results of measurements of the dielectric susceptibility for samples with $x$ close to $x_c$. The dielectric susceptibility ε probes the electrical dipole response of the system and typical results are shown in Fig. 3. We find that for samples measured close to $T_N$ the dielectric function exhibits a frequency dependent peak at a temperature $T_m(f)$ in the real part ε'($T$), which arises from magneto-electric coupling, presumably via striction, plus a peak in the imaginary part ε"($T$) at slightly lower temperatures. Spin-phonon coupling in $BaFe_{12}O_{19}$ has been reported previously from Raman spectroscopy [44-46], and detailed dynamics given by Fontcuberta's group [47-49]. A model for dielectric loss at Néel temperatures has been given by Pirč et al [50], and their graph of ε'($T$) and ε"($T$) for low-frequency probes is given in Fig. 6 of Ref. [50] for realistic parameters, assuming a magneto-electric interaction through striction. The data indicate that polarized clusters form around $T_N$ with glassy dynamics which freeze at $T_m(f)$. Such a state is known as a ferroelectric relaxor or electric-dipole glass [51, 52]. Relaxor dynamics are characterized by a broad distribution of relaxation times, and the freezing process at $T_m(f)$ is associated with the divergence of the longest relaxation time. The present data satisfy a Vogel-Fulcher relationship as in Fig. 3c with frequencies $f$ from 100 Hz to 1 MHz. The glass freezing temperature $T_f = T_m(f\rightarrow 0)$ is found to be finite for samples with $T_N > 0$ and suppressed below zero for samples with paramagnetic ground states $x > x_c$. The electrically polarized clusters which form the glass state are likely supported within the magnetic Fe rich clusters. These magnetic clusters are of diminishing size as the Ga concentration is increased suppressing $T_m(f)$. Since the glass freezing temperature can be tuned through zero with frequency and composition, future studies may involve models of a quantum dipole glass (quantum relaxor) analogous to those studied in spin systems. Both the ferroelectric-glass and the magnetic clusters will contribute to the heat capacity which is likely to be the origin of the non-cubic temperature dependence as observed over any abscissa range in Fig. 2a.

In summary, randomly substituting nonmagnetic Ga ions for magnetic Fe ions in the ferrimagnetic hexagonal ferrite $PbFe_{12-x}Ga_xO_{19}$ suppresses the Néel temperature to zero at a critical composition $x_c$ close to the magnetic ion percolation threshold as calculated for the hexaferrite structure. The phase transition boundary features a $T_N \sim (1 - x/x_c)^{2/3}$ dependence over a wide range of $T$ and $x$. This has not been observed experimentally in other percolation systems, and the origin of this behavior is currently unexplained by theory. Close to $x_c$ the system develops magnetic clusters and an electric-dipole glass with Vogel-Fulcher behaviour. The magneto-electric effect raises the exciting possibility of manipulating the low temperature magnetic phases by electric fields (voltage gates) and the electric-dipole glass by magnetic fields. Future experimental and theoretical studies are likely to be key in elucidating the exotic spin and electric-dipole states and their applications expected to arise close to zero temperature phase transitions in hexaferrites.

**METHODS**

M-type hexaferrite samples were prepared by the flux method. The raw powders of $PbCO_3$, $Fe_2O_3$, $Ga_2O_3$ and fluxing agent $Na_2CO_3$ were weighed in the correct molar ratio and mixed well. The mixed raw powder was put in a platinum crucible and heated to 1250°C for 24 hours in air, then cooled down to 1100°C at a rate of 3°C/min and finally quenched to room temperature. The samples (ca. 2 mm across) were characterized by x-ray diffraction at room temperature using a Rigaku X-ray diffractometer.   Heat capacity was measured as a function of temperature using the relaxation technique on a 5mg sample.   The low temperature DC magnetisation was measured using a SQUID magnetometer up to fields as high as 5 T.   The dielectric measurements were carried out in a liquid-cryogen free cryostat at temperatures as low as 6 K. Silver paste was painted on the surfaces of a thin plate of each crystal and an Andeen-Hagerling, Agilent 4980A and QuadTech LCR instruments were used to measure the dielectric susceptibility at frequencies in the range 100 Hz to 1 MHz.

**ACKNOWLEDGEMENTS**

We would like to thank M. Continentino and G. G. Lonzarich for useful help and discussions.

**CONTRIBUTIONS**

S.E.R. and J.F.S. designed the project.   S.E.R., A.T.J., W.G., J.O., F.D.M., N.L., E.B.S., B.E.W., T.V. and J.F.S. collected the data and interpreted the results.   B.E.W. prepared the samples. T.V. performed the numerical calculations.   S.E.R., T.V. and J.F.S. analysed the data and wrote the paper.

**FUNDING**

S.E.R. and E.B.S. acknowledge support from a CONFAP Newton grant. T.V. acknowledges support from the NSF under Grant No. DMR-1506152.

**FIGURES**

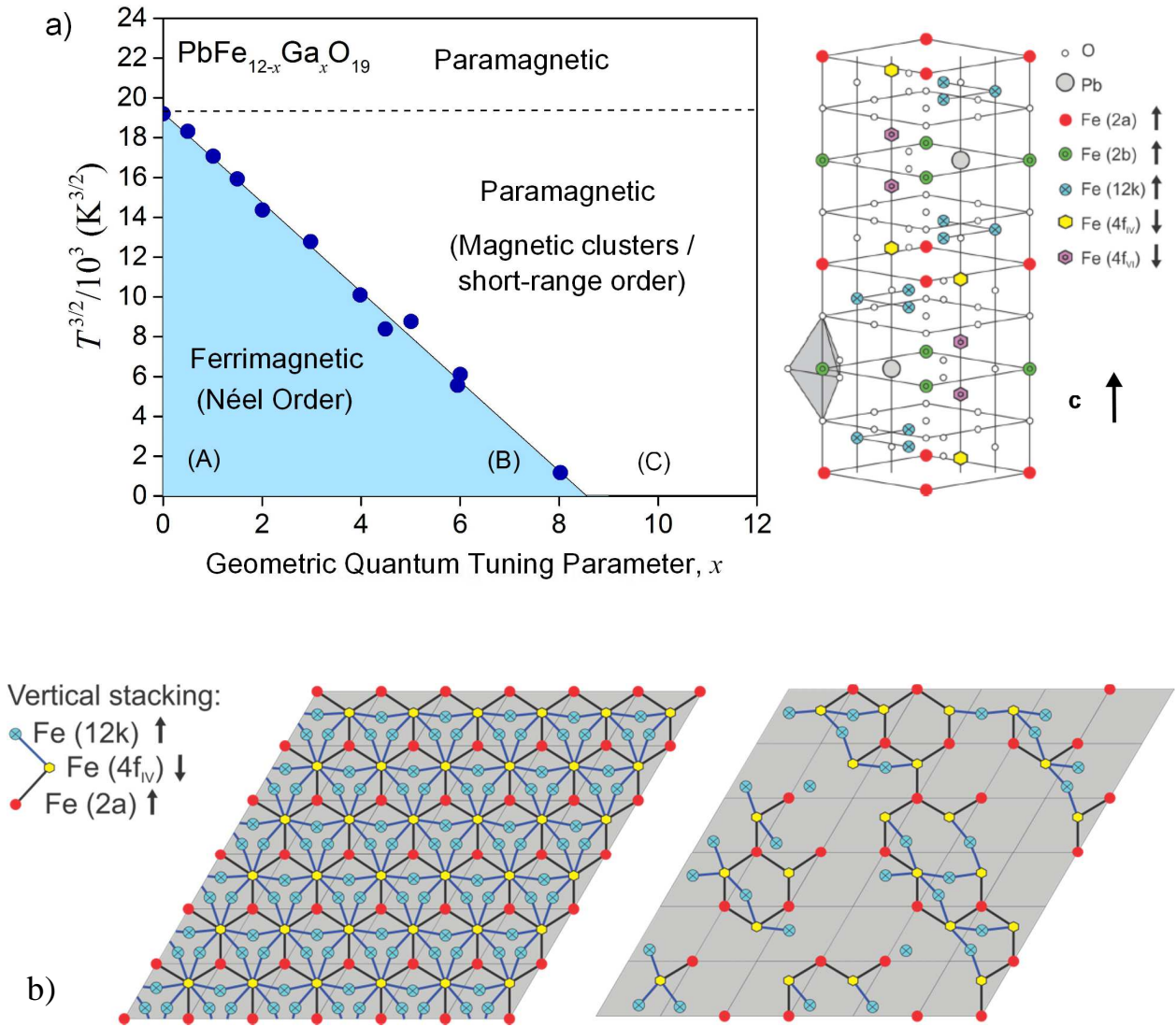

**Figure 1 –Magnetic phase diagram and crystal structure of PbFe$_{12-x}$Ga$_x$O$_{19}$.** The Néel temperature, $T_N \approx 718K$, in PbFe$_{12}$O$_{19}$ separating paramagnetic and ferrimagnetic phases is suppressed via non-magnetic Ga substitution and tuned through a geometrically driven percolation phase transition located at $T = 0K$ and $x = x_c \approx 8.6$ as shown in (a). $x = 8.6$ is close to the calculated percolation threshold $x = 8.85$ referred to in the main text for the hexaferrite structure. The right inset shows a double unit-cell of PbFe$_{12}$O$_{19}$ as explained more fully in the main text with the crystallographic $c$ direction indicated by the arrow. Values of $T_N$ were determined by Mössbauer

and magnetisation measurements [6]. The value of $T_N$ as a function of Ga $x$ in the related materials BaFe$_{12-x}$Ga$_x$O$_{19}$ and SrFe$_{12-x}$Ga$_x$O$_{19}$ differ from those shown above for PbFe$_{12-x}$Ga$_x$O$_{19}$ by only a few per cent. The main figure shows $T_N^{3/2}$ (blue dots) plotted against $x$ and the straight line is a best fit to the data with an equation of the form $T_N/(718 \text{ K}) = (1 - x/x_c)^\phi$ with critical Ga concentration $x_c = 8.56$, and the power-law exponent determined as $\phi = 0.67 \pm 0.02$, i.e. 2/3. The region labelled (A) is where uniaxial quantum critical ferroelectric fluctuations have recently been reported in BaFe$_{12}$O$_{19}$ and SrFe$_{12}$O$_{19}$ [3, 9]. The regions labelled (B) and (C) are where an electric-dipole glass state (ferroelectric relaxor) is observed, induced by magneto-electric coupling as explained in the main text and later figures. The dashed line separates the classical paramagnetic phase and the paramagnetic phase composed of disconnected clusters of ferrimagnetic order. The region labelled (C) is where one might expect to search for exotic spin and thermodynamic states [27, 28]. In (b) the left image shows a projection into the a-b plane of the relevant magnetic ions (one spinel block shown) used in the percolation calculation explained in the main text for the un-doped parent compound PbFe$_{12}$O$_{19}$. The right image shows a percolating magnetic cluster under the conditions of magnetic dilution in PbFe$_{12-x}$Ga$_x$O$_{19}$ with $x$ close to $x_c$.

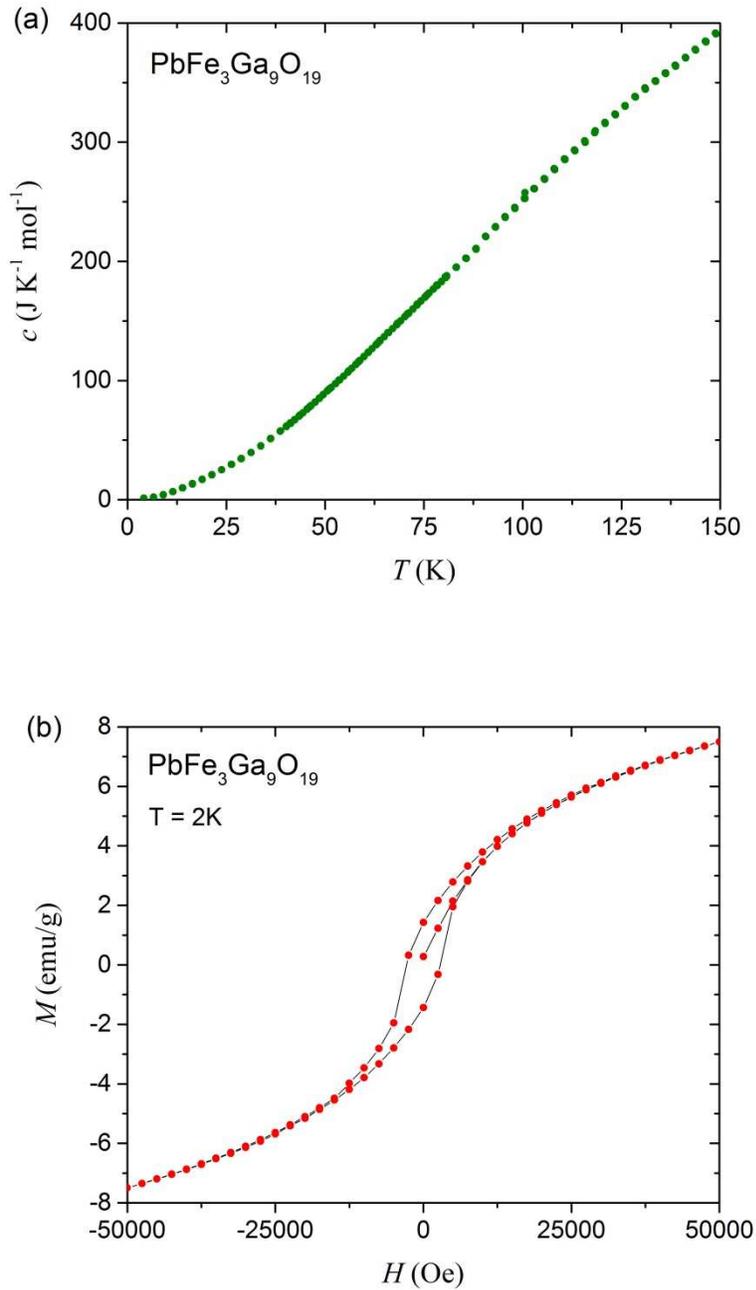

**Figure 2 – Thermal and magnetization measurements in PbFe₃Ga₉O₁₉ ($x = 9$).** The heat capacity as a function of temperature in (a) demonstrates along with Mössbauer experiments [6] the absence of a bulk phase transition and thus no long-range order in a sample with $x > x_c$. The weak hysteresis measured at 2K in (b) indicates the contribution to the uniform magnetization from 'rare regions' - small disconnected ferrimagnetic clusters - in the paramagnetic phase at low temperatures. Close to the zero temperature percolation phase transition the magnetic clusters are randomly distributed in space but perfectly correlated in time leading to the possibility of singular thermodynamic functions over a range of tuning parameter variables [20, 28].

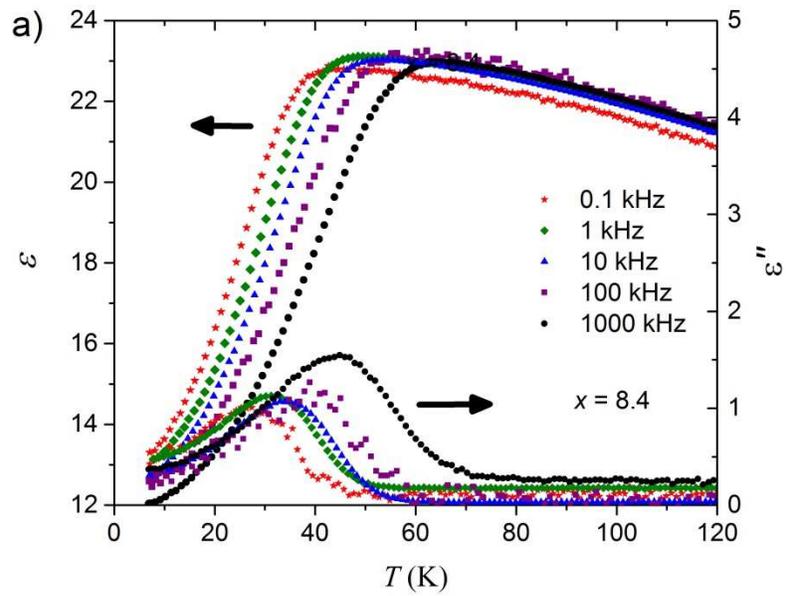

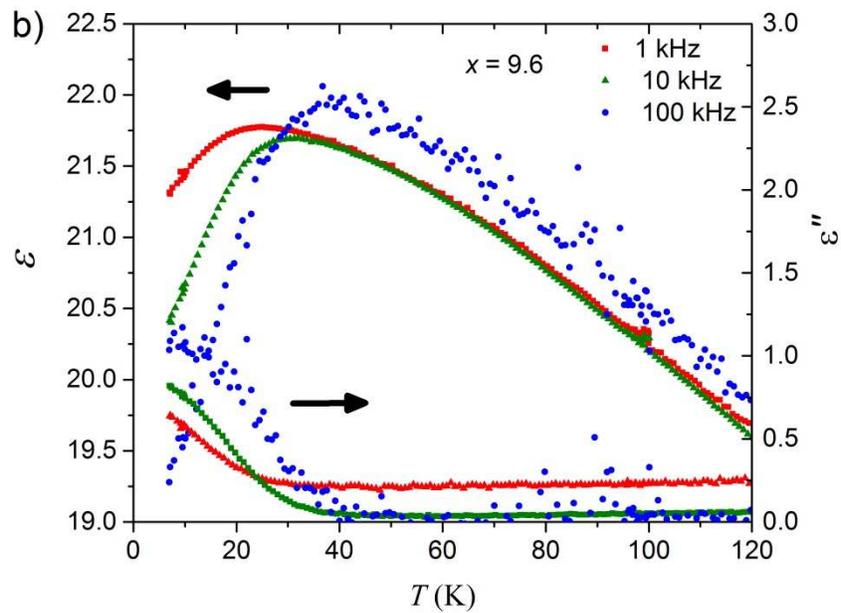

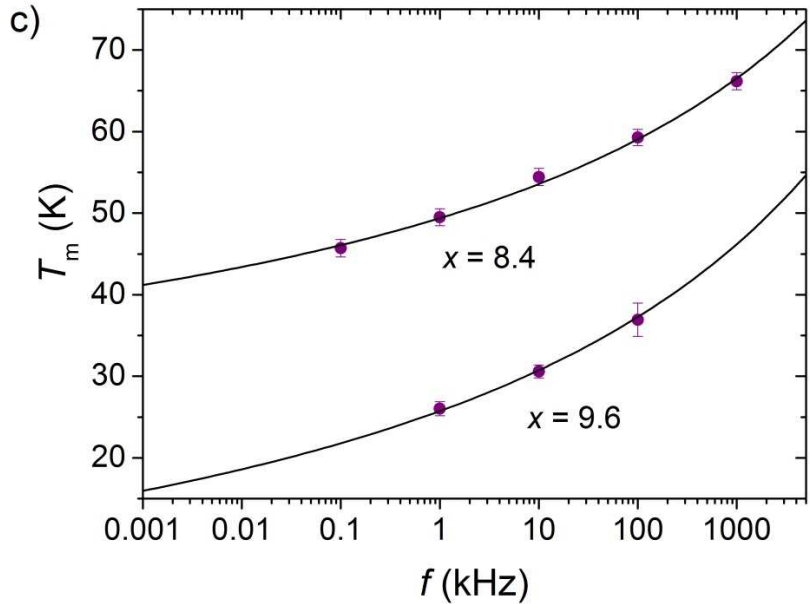

**Figure 3 - Real and imaginary parts of the dielectric constant (a, b) and Vogel-Fulcher plots (c), for PbFe$_{12-x}$Ga$_x$O$_{19}$ showing dipole-glass behaviour.** Figures (a) and (b) show the real $\varepsilon$ and imaginary parts $\varepsilon''$ of the dielectric constant measured at different frequencies plotted against temperature $T$ for samples of PbFe$_{12-x}$Ga$_x$O$_{19}$ with $x = 8.4$ and $x = 9.6$ respecitvely. A Vogel-Fulcher fit to the data of the peak temperature $T_m$ versus measurement frequency $f$ is shown for the same two samples in (c). The Vogel-Fulcher equation is of the form $f = f_0 exp\left(-T_a/(T_m - T_f)\right)$ where the constant $T_a$ is the activation temperature scale and $f_0$ is a characteristic frequency. The frequency dependent variable $T_m(f)$ is defined as the temperature at which $\varepsilon'$ has a peak in $T$ as in the examples shown in (a). The constant $T_f$ is the freezing temperature in the zero frequency limit. For $x = 9.6$ the fitting parameters were $T_a = 730$ K, $T_f = -11.7$ K and $f_0 = 3.02 \times 10^{11}$ Hz and for $x = 8.4$ they were $T_a = 611$ K, $T_f = 18.1$ K and $f_0 = 3.00 \times 10^{11}$ Hz.

# REFERENCES

[1]   R. C. Pullar, Prog. Mater. Sci. **57**, 1191 (2012).
[2]   R. C. Pullar, in *Mesoscopic Phenomena in Multifunctional Materials: Synthesis, Characterization, Modeling and Applications*, edited by A. Saxena, and A. Planes (Springer-Verlag Berlin, Berlin, 2014), pp. 159.
[3]   S. E. Rowley *et al.*, Scientific Reports **6,** 25724 (2016).
[4]   S. P. Shen *et al.*, Nat. Commun. **7,** 10569 (2016).
[5]   E. Lieb, and D. Mattis, Journal of Mathematical Physics **3**, 749 (1962).
[6]   G. Albanese *et al.*, J. Mater. Sci. **37**, 3759 (2002).
[7]   C. M. Fang *et al.*, Journal of Physics-Condensed Matter **15**, 6229 (2003).
[8]   A. S. Mikheykin *et al.*, European Physical Journal B **87**, 232 (2014).
[9]   S. P. Shen *et al.*, Phys. Rev. B **90,** 180404(R) (2014).
[10]   A. Grill, and F. Haberey, Appl Phys **3**, 131 (1974).
[11]   A. Isalgue *et al.*, Appl Phys a-Mater **39**, 221 (1986).
[12]   C. J. Wu *et al.*, Scientific Reports **6**, 36200 (2016).
[13]   M. E. J. Newman, and R. M. Ziff, Physical Review E **64**, 016706 (2001).
[14]   D. Stauffer, and A. Aharony, *Introduction to percolation theory* (Taylor & Francis, London, 1994).
[15]   K. J. Schrenk, N. A. M. Araujo, and H. J. Herrmann, Physical Review E **87**, 032123 (2013).
[16]   R. B. Griffiths, Phys. Rev. Lett. **23**, 17 (1969).
[17]   T. Vojta, J Phys A-Math Gen **39**, R143 (2006).
[18]   O. P. Vajk *et al.*, Science **295**, 1691 (2002).
[19]   R. J. Birgeneau *et al.*, Journal of Statistical Physics **34**, 817 (1984).
[20]   T. Senthil, and S. Sachdev, Phys. Rev. Lett. **77**, 5292 (1996).
[21]   A. Coniglio, Phys. Rev. Lett. **46**, 250 (1981).
[22]   S. Sachdev, Nat. Phys. **4**, 173 (2008).
[23]   S. Sachdev, *Quantum phase transitions* (Cambridge University Press, Cambridge ; New York, 2011), pp. xviii.
[24]   X.-G. Wen, *Quantum field theory of many-body systems : from the origin of sound to an origin of light and electrons* (Oxford University Press, Oxford ; New York, 2004), pp. xiii.
[25]   S. Sachdev, and T. Senthil, Ann. Phys. **251**, 76 (1996).
[26]   C. Balz *et al.*, Nat. Phys. **12**, 942 (2016).
[27]   T. Vojta, AIP Conf. Proc. **1550**, 188 (2013).
[28]   T. Vojta, and J. Schmalian, Phys. Rev. Lett. **95**, 237206 (2005).
[29]   A. B. Rechester, Sov. Phys. JETP **33**, 423 (1971).
[30]   D. E. Khmelnitskii, and V. L. Shneerson, Sov. Phys. - Solid State **13**, 687 (1971).
[31]   D. E. Khmelnitskii, and V. L. Shneerson, Sov. Phys. JETP **37**, 164 (1973).
[32]   K. K. Murata, and S. Doniach, Phys. Rev. Lett. **29**, 285 (1972).
[33]   T. Moriya, and A. Kawabata, J. Phys. Soc. Jpn. **34**, 639 (1973).
[34]   T. Moriya, and A. Kawabata, J. Phys. Soc. Jpn. **35**, 669 (1973).
[35]   Ramakris.Tv, Phys. Rev. B **10**, 4014 (1974).
[36]   I. E. Dzyaloshinskii, and P. S. Kondratenko, Zhurnal Eksp. Teor. Fiz. **70**, 1987 (1976).
[37]   J. A. Hertz, Phys. Rev. B **14**, 1165 (1976).
[38]   G. G. Lonzarich, and L. Taillefer, J Phys C Solid State **18**, 4339 (1985).

[39] A. J. Millis, Physical Review B **48**, 7183 (1993).

[40] S. E. Rowley *et al.*, Nature Physics **10**, 367 (2014).

[41] M. A. Continentino, *Quantam scaling in many-body systems* (World Scientific, Singapore ; London, 2001).

[42] G. G. Lonzarich (Ch. 6), and M. Springford (Editor), *Electron : a centenary volume* (Cambridge University Press, Cambridge, 1997), pp. xii.

[43] A. B. Harris, Journal of Physics C-Solid State Physics **7**, 1671 (1974).

[44] X. B. Chen *et al.*, Chin. Phys. Lett. **29,** 126103 (2012).

[45] T. M. H. Nguyen *et al.*, Journal of Raman Spectroscopy **43**, 2020 (2012).

[46] X. B. Chen *et al.*, J. Appl. Phys. **114,** 013912 (2013).

[47] J. Muller, and A. Collomb, J. Magn. Magn. Mater. **103**, 194 (1992).

[48] J. Fontcuberta, and X. Obradors, Journal of Physics C-Solid State Physics **21**, 2335 (1988).

[49] X. Obradors *et al.*, J. Solid State Chem. **56**, 171 (1985).

[50] R. Pirc, R. Blinc, and J. F. Scott, Phys. Rev. B **79**, 214114 (2009).

[51] G. A. Samara, in *Solid State Physics, Vol 56*, edited by H. Ehrenreich, and F. Spaepen (Elsevier Academic Press Inc, San Diego, 2001), pp. 239.

[52] R. Pirc, and R. Blinc, Phys. Rev. B **76,** 020101(R) (2007).